УДК 681.5

# РОБАСТНОЕ ТРАЕКТОРНОЕ УПРАВЛЕНИЕ КВАДРОКОПТЕРОМ ПО ВЫХОДУ НА ОСНОВЕ ГЕОМЕТРИЧЕСКОГО ПОДХОДА

О.И. Борисов, М.А. Каканов, А.Ю. Живицкий, А.А. Пыркин

Университет ИТМО, 197101, Санкт-Петербург, Россия
E-mail: borisov@itmo.ru

В работе решена задача траекторного управления квадрокоптером с неизмеримыми углами тангажа и крена на основе геометрического подхода с применением модифицированного расширенного наблюдателя и внутренней модели. Предлагаемый подход позволяет обеспечить движение квадрокоптера в горизонтальной плоскости по траектории, заданной в форме периодической (синусоидальной) функции или полиномиальной функции второго порядка, с полуглобальной асимптотической сходимостью ошибок слежения к нулю.

*Ключевые слова*: *траекторное управление, геометрический подход, робастное управление, наблюдатель с высоким коэффициентом усиления, внутренняя модель, квадрокоптер.*

**Введение.** Геометрической подход представляет собой комплекс методов анализа и синтеза систем управления, в основе которых лежит исследование динамических свойств систем как геометрических характеристик пространств состояний и их подпространств, некоторые их которых, в частности, могут являться инвариантными при замены координат, что, как отмечается в [1], в ряде случаев позволяет упростить решение той или иной задачи управления. Последнее особенно актуально для нелинейных систем, где задача синтеза управления в общем виде является нетривиальной, а геометрический подход позволяет формализовать процедуру синтеза управления путем замены координат и приведения динамической модели объекта к так называемой нормальной форме, к которой применима линеаризация по обратной связи, а также синтез робастного закона управления на базе оценок неопределенных функций, полученных с помощью наблюдателя с высоким коэффициентом усиления. Методы геометрического подхода для нелинейных систем приведены в книгах [2-4], однако развитие методов геометрического подхода сохраняет свою актуальность, в том числе в части расширения класса нелинейных систем и задач управления, для которых они применимым. В работе [5] предложен синтез модифицированного расширенного наблюдателя (enhanced extended observer) на базе расширенной леммы Даяванца. В сравнении со стандартным расширенным наблюдателем, предложенным в [6], модифицированная версия позволяет охватить более широкий класс нелинейных систем, нормальная форма которых характеризуются нестационарным коэффициентом усиления. К такому классу систем, в частности, относится динамическая модель квадрокоптера, у которого, как будет показано ниже, нестационарный коэффициент усиления в нормальной форме зависит от управляющего воздействия, соответствующего вертикальному движению.

Настоящая статья представляет собой расширение результатов работы [7], в которой показано применение вышеупомянутого модифицированного расширенного наблюдателя для решения задачи удержания заданного положения квадрокоптера с неизмеряемыми углами крена и тангажа. Усиление результатов настоящей статьи связано с решением задачи траекторного управления квадрокоптером в горизонтальной плоскости путем применения внутренней модели для компенсации в установившемся режиме действия внешних воздействий, представляющих собой задающие сигналы, описываемые генератором траектории, и, как следствие, обеспечения полуглобальной асимптотической сходимости ошибок слежения к нулю.

**Постановка задачи**. Рассмотрим динамическую модель квадрокоптера, полученную после преобразований, произведенных в работе [7],

$$\begin{aligned} \ddot{x} &= (g + u_0)\sin\theta\cos\psi, \\ \ddot{y} &= -(g + u_0)\sin\psi, \\ \ddot{z} &= (g + u_0)\cos\theta\cos\psi - g, \\ \ddot{\theta} &= u_1, \\ \ddot{\psi} &= u_2, \end{aligned} \qquad (1)$$

где $x, y, z \in \mathbb{R}$ — декартовы координаты, доступные для измерения, $\theta, \psi \in \mathbb{S} := [0, 2\pi)$ — соответственно углы тангажа и крена, предполагающиеся неизмеримыми, $u_0, u_1, u_2$ — управляющие сигналы, $g = 9{,}81 \text{ м/с}^2$ — ускорение свободного падения.

Следуя [7], примем, что задача стабилизации квадрокоптера на заданной высоте $z^* = \text{const}$ выполнена с помощью ПД-регулятора с насыщением

$$u_0 = \text{sat}_\ell[-r_0\tilde{z} - r_1\dot{z}], \quad 0 < \ell < g, \qquad (2)$$

который при соответствующем выборе параметров $r_0, r_1$ и выполнении условия $|\cos\theta\cos\psi| \geq \frac{g}{g+\ell}$ обеспечивает устойчивость вертикального движения квадрокоптера по входу-состоянию и сходимость ошибки стабилизации $\tilde{z} = z(t) - z^*$ к нулю.

Требуется синтезировать робастный регулятор по выходу на основе геометрического подхода, обеспечивающий траекторное управление квадрокоптером в горизонтальной плоскости и выполнение целевых условий

$$\begin{aligned} \lim_{t\to\infty}|\tilde{x}(t)| &= 0, \quad \tilde{x}(t) = x(t) - x^*(t), \\ \lim_{t\to\infty}|\tilde{y}(t)| &= 0, \quad \tilde{y}(t) = y(t) - y^*(t), \end{aligned} \qquad (3)$$

где $x^*(t)$ и $y^*(t)$ — задающие сигналы, определяющие желаемую траекторию движения в горизонтальной плоскости и описываемые генератором вида

$$\begin{aligned} \begin{pmatrix} \dot{w}_1 \\ \dot{w}_2 \end{pmatrix} &= \begin{pmatrix} S_1 & 0 \\ 0 & S_2 \end{pmatrix}\begin{pmatrix} w_1 \\ w_2 \end{pmatrix}, \\ \begin{pmatrix} x^* \\ y^* \end{pmatrix} &= \begin{pmatrix} H_1 & 0 \\ 0 & H_2 \end{pmatrix}\begin{pmatrix} w_1 \\ w_2 \end{pmatrix}, \end{aligned} \qquad (4)$$

где

$$S_i = \begin{pmatrix} 0 & 1 & 0 \\ 0 & 0 & 1 \\ 0 & -\rho_i & 0 \end{pmatrix}, \quad w_i = \begin{pmatrix} w_{11} \\ w_{i2} \\ w_{i3} \end{pmatrix}, \quad H_i = \begin{pmatrix} 1 & 0 & 0 \end{pmatrix}, \quad i = \{1,2\},$$

где $\rho_i \geq 0$ — некоторый заданный параметр.

*Замечание 1.* В зависимости от выбора параметра $\rho_i$ и начальных условий $w_i(0)$ генератор (4) позволяет вырабатывать сигнал в синусоидальной или полиномиальной форме второго порядка.

Действительно, при $\rho_i = \omega^2$ и $w_i(0) = (C_i + A_i\sin\varphi_i \quad A_i\omega_i\cos\varphi_i \quad -A_i\omega_i^2\sin\varphi_i)^T$ сигнал принимает вид

$$\begin{pmatrix} x^*(t) \\ y^*(t) \end{pmatrix} = \begin{pmatrix} C_1 + A_1\sin(\omega_1 t + \varphi_1) \\ C_2 + A_2\sin(\omega_2 t + \varphi_2) \end{pmatrix},$$

а при $\rho_i = 0$ и $w_i(0) = (C_{i0} \quad C_{i1} \quad 2C_{i2})^T$ сигнал принимает вид

$$\begin{pmatrix} x^*(t) \\ y^*(t) \end{pmatrix} = \begin{pmatrix} C_{10} + C_{11}t + C_{12}t^2 \\ C_{20} + C_{21}t + C_{22}t^2 \end{pmatrix}. \tag{5}$$

**Основной результат.** Выполним предложенную в работе [7] замену координат

$$\begin{array}{llll} \xi_{11} &=& x, & \xi_{21} &=& y, \\ \xi_{12} &=& \dot{x}, & \xi_{22} &=& \dot{y}, \\ \xi_{13} &=& g\sin\theta\cos\psi, & \xi_{23} &=& -g\sin\psi, \\ \xi_{14} &=& g\dot{\theta}\cos\theta\cos\psi - g\dot{\psi}\sin\theta\sin\psi, & \xi_{24} &=& -g\dot{\psi}\cos\psi, \end{array} \tag{6}$$

являющуюся обратимой при $|\theta| < \pi/2$ и $|\psi| < \pi/2$, что соответствует номинальному режиму полета квадрокоптера, и получим систему в нормальной форме

$$\begin{array}{llll} \dot{\xi}_{11} &=& \xi_{12}, & \dot{\xi}_{21} &=& \xi_{22}, \\ \dot{\xi}_{12} &=& \beta(t)\xi_{13}, & \dot{\xi}_{22} &=& \beta(t)\xi_{23}, \\ \dot{\xi}_{13} &=& \xi_{14}, & \dot{\xi}_{23} &=& \xi_{24}, \\ \dot{\xi}_{14} &=& q_1(\theta,\dot{\theta},\psi,\dot{\psi}) + b_1(\theta,\psi)\begin{pmatrix} u_1 \\ u_2 \end{pmatrix}, & \dot{\xi}_{24} &=& q_2(\psi,\dot{\psi}) + b_2(\psi)\begin{pmatrix} u_1 \\ u_2 \end{pmatrix}, \end{array}$$

где $q_1(\theta,\dot{\theta},\psi,\dot{\psi})$, $b_1(\theta,\psi)$, $q_2(\psi,\dot{\psi})$, $b_2(\psi)$ – нелинейные неопределенные функции вида

$$\begin{aligned}
q_1(\theta,\dot{\theta},\psi,\dot{\psi}) &= g\ddot{\theta}\cos\theta\cos\psi - g\ddot{\psi}\sin\theta\sin\psi - 2g\dot{\psi}\dot{\theta}\cos\theta\sin\psi - g[\dot{\theta}^2 + \dot{\psi}^2]\sin\theta\cos\psi, \\
b_1(\theta,\psi) &= (g\cos\theta\cos\psi \quad -g\sin\theta\sin\psi), \\
q_2(\psi,\dot{\psi}) &= g\dot{\psi}^2\sin\psi - g\ddot{\psi}\cos\psi, \\
b_2(\psi) &= (0 \quad -g\cos\psi),
\end{aligned}$$

а также $\beta(t)$ – ограниченная измеримая функция вида

$$\beta(t) = \left(1 + \frac{u_0(t)}{g}\right), \quad 0 < \beta_{min} \leq \beta(t) \leq \beta_{max},$$

где $\beta_{min} = 1 - \frac{\ell}{g}$ и $\beta_{max} = 1 + \frac{\ell}{g}$.

Зададим сигналы ошибок $e_1 = (e_{11} \quad \dots \quad e_{14})^T, e_2 = (e_{21} \quad \dots \quad e_{24})^T$ как

$$\begin{aligned}
e_{11} &= \xi_{11} - w_{11}, & e_{21} &= \xi_{21} - w_{21}, \\
e_{12} &= \xi_{12} - w_{12}, & e_{22} &= \xi_{22} - w_{22}, \\
e_{13} &= \xi_{13} - \frac{1}{\beta(t)}w_{13}, & e_{23} &= \xi_{23} - \frac{1}{\beta(t)}w_{23}, \\
e_{14} &= \xi_{14} + \frac{\dot{\beta}}{\beta^2}w_{13} + \frac{\rho_1}{\beta}w_{12}, & e_{24} &= \xi_{24} + \frac{\dot{\beta}}{\beta^2}w_{23} + \frac{\rho_2}{\beta}w_{22},
\end{aligned}$$

продифференцировав которые получим модели ошибок

$$\begin{aligned}
\dot{e}_{11} &= e_{12}, \\
\dot{e}_{12} &= \beta(t)e_{13}, \\
\dot{e}_{13} &= e_{14}, \\
\dot{e}_{14} &= q_1(\theta,\dot{\theta},\psi,\dot{\psi}) + \left[\frac{\ddot{\beta}}{\beta^2} - \frac{2\dot{\beta}^2}{\beta^3} + \frac{\rho_1}{\beta}\right]w_{13} - \frac{2\rho_1\dot{\beta}}{\beta^2}w_{12} + b_1(\theta,\psi)\begin{pmatrix}u_1\\u_2\end{pmatrix},
\end{aligned} \quad (7)$$

и

$$\begin{aligned}
\dot{e}_{21} &= e_{22}, \\
\dot{e}_{22} &= \beta(t)e_{23}, \\
\dot{e}_{23} &= e_{24}, \\
\dot{e}_{24} &= q_2(\psi,\dot{\psi}) + \left[\frac{\ddot{\beta}}{\beta^2} - \frac{2\dot{\beta}^2}{\beta^3} + \frac{\rho_2}{\beta}\right]w_{23} - \frac{2\rho_2\dot{\beta}}{\beta^2}w_{22} + b_2(\psi)\begin{pmatrix}u_1\\u_2\end{pmatrix}.
\end{aligned} \quad (8)$$

Запишем модель (7), (8) в общем виде как

$$\begin{aligned}
\dot{w} &= s(w), \\
\dot{e} &= f(w,e,u), \\
\begin{pmatrix}\tilde{x}\\\tilde{y}\end{pmatrix} &= h_e(w,e),
\end{aligned} \quad (9)$$

где $w = (w_1 \quad w_2)^T$ — вектор состояния генератора, $e = (e_1 \quad e_2)^T$ — вектор состояния модели ошибки, $u = (u_1 \quad u_2)^T$ – вектор управляющих воздействий, $s(\cdot), f(\cdot), h_e(\cdot)$ — соответствующие функции.

Исследуем поведение системы (9) в установившемся режиме с помощью выражений

$$\frac{\partial \pi}{\partial w} s(w) = f(w, \pi(w), v(w)),$$
$$0 = h_e(w, \pi(w)),$$

где $\pi(w) = (\pi_1(w) \quad \pi_2(w))^T$ — значение вектора состояния объекта в установившемся режиме, $v(w) = (v(w_1) \quad v_2(w))^T$ — значение вектора управления в установившемся режиме.

Заметим, что в установившемся режиме справедливы соотношения

$$q_1(0,0,0,0) = 0, \quad b_1(0,0) = (g \quad 0), \quad q_2(0,0) = 0, \quad b_2(0) = (0 \quad -g),$$
$$u_0(\infty) = 0, \quad \beta(\infty) = 1, \quad \dot{\beta}(\infty) = \ddot{\beta}(\infty) = 0,$$

с учетом которых найдем пару $\pi(2), v(w)$ для решения задачи управления по выходу системой (7), (8)

$$\pi(w) = (\pi_1(w) \quad \pi_2(w))^T = (0 \quad 0 \quad 0 \quad 0 \quad 0 \quad 0 \quad 0 \quad 0)^T,$$
$$v(w) = (v_1(w) \quad v_2(w))^T = \left(-\frac{\rho_1}{g} w_{13} \quad \frac{\rho_2}{g} w_{23}\right)^T \coloneqq \Psi w. \quad (10)$$

Для компенсации внешних сигналов $w$ и, как следствие, обеспечения сходимости ошибок слежения к нулю сформируем внутреннюю модель вида

$$\dot{\eta} = F\eta + G[\Gamma\eta + \bar{u}],$$
$$u = \Gamma\eta + \bar{u}, \quad (11)$$

где $\bar{u}$ — управляющие воздействие, которое будет определено позднее, $F = \text{diag}(F_1, F_2)$ — гурвицева матрица, образующая вместе с $G = \text{diag}(G_1, G_2), G_1 = G_2 = (0 \quad 0 \quad ... \quad 0 \quad 1)^T$ управляему пару $(F, G)$, матрица $\Gamma = \text{diag}(\Gamma_1, \Gamma_2)$ такая, что собственные числа матрицы $\Phi = F + G\Gamma$ совпадают с собственными числами матрицы $S = \text{diag}(S_1, S_2)$.

Известно [4, 8], что существует матрица $\Sigma$ такая, что

$$\begin{aligned}\Sigma S &= (F + G\Gamma)\Sigma, \\ \Psi &= \Gamma\Sigma.\end{aligned}$$

*Замечание 2.* Если параметр $\rho_i$ генератора (4) равен нулю, то, как следует из выражения (10), значение управляющего воздействия $v_i(w)$ в установившемся режиме также равно нулю $\rho_i = 0 \Rightarrow v_i(w) = 0$, что говорит о том, что задача слежения за квадратичным задающим сигналом вида (5) не требует построения внутренней модели (11).

Объединим внутреннюю модель с объектом управления и получим агрегированную систему вида

$$\begin{aligned}\dot{\eta} &= F\eta + G[\Gamma\eta + \bar{u}], \\ \dot{e}_{i1} &= e_{i2}, \\ \dot{e}_{i2} &= \beta(t)e_{i3}, \\ \dot{e}_{i3} &= e_{i4}, \\ \dot{e}_{i4} &= q_i(\cdot) + \left[\frac{\ddot{\beta}}{\beta^2} - \frac{2\dot{\beta}^2}{\beta^3} + \frac{\rho_i}{\beta}\right]w_{i3} - \frac{2\rho_i\dot{\beta}}{\beta^2}w_{i2} + b_i(\cdot)[\Gamma\eta + \bar{u}].\end{aligned} \quad (12)$$

Введем новую переменную

$$\tilde{\eta} = \eta - \Sigma w,$$

производная которой равна

$$\dot{\tilde{\eta}} = F\tilde{\eta} + G[\Gamma\tilde{\eta} + \bar{u}],$$

с учетом чего система (12) принимает вид

$$\begin{aligned}\dot{\tilde{\eta}} &= F\tilde{\eta} + G[\Gamma\tilde{\eta} + \bar{u}], \\ \dot{e}_{i1} &= e_{i2}, \\ \dot{e}_{i2} &= \beta(t)e_{i3}, \\ \dot{e}_{i3} &= e_{i4}, \\ \dot{e}_{i4} &= \tilde{q}_i(\cdot) + b_i(\cdot)[\Gamma\tilde{\eta} + \bar{u}],\end{aligned} \quad (13)$$

где функция

$$\tilde{q}_i(\cdot) = q_i(\cdot) + \left[\frac{\ddot{\beta}}{\beta^2} - \frac{2\dot{\beta}^2}{\beta^3} + \frac{\rho_i}{\beta}\right]w_{i3} - \frac{2\rho_i\dot{\beta}}{\beta^2}w_{i2} + b_i(\cdot)\Psi w$$

в положении равновесия $e_i = (0 \quad 0 \quad 0 \quad 0)^T$ равна нулю по построению $\Psi$.

Для системы (13) выберем номинальный линеаризующий закон управления вида

$$\bar{u} = -\Gamma\tilde{\eta} + B^{-1}(\theta,\psi)\begin{pmatrix} -\tilde{q}_1(\theta,\dot{\theta},\psi,\dot{\psi},w,\beta,\dot{\beta},\ddot{\beta}) + Ke_1 \\ -\tilde{q}_2(\psi,\dot{\psi},w,\beta,\dot{\beta},\ddot{\beta}) + Ke_2 \end{pmatrix}, \qquad (14)$$

где

$$B(\theta,\psi) = \begin{pmatrix} b_1(\theta,\psi) \\ b_2(\psi) \end{pmatrix} = \begin{pmatrix} g\cos\theta\cos\psi & -g\sin\theta\sin\psi \\ 0 & -g\cos\psi \end{pmatrix}, \quad K = \begin{pmatrix} k_1 & k_2 & k_3 & k_4 \end{pmatrix}.$$

Подставляя закон управления (14) в (13), получим замкнутую систему вида

$$\begin{aligned}
\dot{\tilde{\eta}} &= F\tilde{\eta} + GB^{-1}(\theta,\psi)\begin{pmatrix} -\tilde{q}_1(\theta,\dot{\theta},\psi,\dot{\psi},w,\beta,\dot{\beta},\ddot{\beta}) + Ke_1 \\ -\tilde{q}_2(\psi,\dot{\psi},w,\beta,\dot{\beta},\ddot{\beta}) + Ke_2 \end{pmatrix}, \\
\dot{e}_{i1} &= e_{i2}, \\
\dot{e}_{i2} &= \beta(t)e_{i3}, \\
\dot{e}_{i3} &= e_{i4}, \\
\dot{e}_{i4} &= Ke_i,
\end{aligned} \qquad (15)$$

где вектор $e_i$ экспоненциально сходится к нулю в силу соответствующего выбора вектора $K$ (см. работу [7]), а подсистема, описываемая первым выражением из (15), является устойчивой по входу-состоянию. Таким образом, можно сделать вывод, что вся система (15) является глобально асимптотически устойчивой.

Однако, поскольку закон управления (14) содержит неопределенности, он не является реализуемым. Для построения реализуемого закона управления воспользуемся так называемым подходом модифицированного расширенного наблюдателя, предложенного в [5].

Построим робастный закон управления вида

$$\bar{u} = \text{sat}_\nu\left[\bar{B}^{-1}\begin{pmatrix} -\sigma_1 + K\hat{e}_1 \\ -\sigma_2 + K\hat{e}_2 \end{pmatrix}\right], \qquad (16)$$

где $\text{sat}_\nu$ — гладкая функция насыщения с пределом $\nu$, $\bar{B} = \begin{pmatrix} \bar{b}_1 \\ \bar{b}_2 \end{pmatrix}$ — обратимая матрица такая, что

$$\|[B(\theta,\psi) - \bar{B}]\bar{B}^{-1}\|_1 < 1,$$

а $\hat{e}_i = \begin{pmatrix} e_{i1} & e_{i2} & e_{i3} & e_{i4} \end{pmatrix}$ и $\sigma_i$ при $i = \{1,2\}$ представляют собой состояния модифицированных расширенных наблюдателей вида

$$\begin{aligned}
\dot{\hat{e}}_{i1} &= \hat{e}_{i2} + \kappa a_4(\tilde{x} - \hat{e}_{i1}), \\
\dot{\hat{e}}_{i2} &= \beta(t)\hat{e}_{i3} + \kappa^2 a_3(\tilde{x} - \hat{e}_{i1}), \\
\dot{\hat{e}}_{i3} &= \hat{e}_{i4} + \kappa^3 a_2(\tilde{x} - \hat{e}_{i1}), \\
\dot{\hat{e}}_{i4} &= \sigma_i + \bar{b}_i \bar{u} + \kappa^4 a_1(\tilde{x} - \hat{e}_{i1}), \\
\dot{\sigma}_i &= \kappa^5 a_5(\tilde{x} - \hat{e}_{i1}).
\end{aligned} \qquad (17)$$

Робастный закон управления (11), (16), (17) обеспечивает в системе свойства полуглобальной асимптотической устойчивости, что означает выполнение цели управления (3) для ограниченного множества начальных условий. Формализуем результат в виде утверждения.

*Утверждение 1.* Для заданной траектория движения в горизонтальной плоскости $(x^*, y^*)$, вырабатываемой генератором (4), и ограниченного множества начальных условий $R$ такого, что $\| e(0) \| \leq R$, $\| (\hat{e}(0), \sigma(0)) \| \leq R$, существует предел насыщения $\nu$, параметры $a_0, a_1, a_2, a_3, a_4$ и значение $\kappa^*$ такие, что при $\kappa > \kappa^*$ все траектории системы (7), (8), (11), (16), (17) ограничены и сигналы ошибок слежения $\tilde{x}, \tilde{y}$ асимптотически сходятся к нулю, что означает выполнение цели управления (3).

**Моделирование.** В настоящем разделе рассмотрим моделирование решения задачи траекторного управления квадрокоптером с использованием предложенного подхода.

Выберем высоту полета $z^* = 0{,}5$ и рассмотрим две формы заданной траектории в горизонтальной плоскости:

1) периодический сигнал

$$\begin{pmatrix} x^*(t) \\ y^*(t) \end{pmatrix} = \begin{pmatrix} 0{,}1 + 0{,}3\sin(0{,}5t + 0{,}7) \\ 0{,}2 + 0{,}4\sin(0{,}6t + 0{,}8) \end{pmatrix}; \qquad (18)$$

2) полиномиальный сигнал

$$\begin{pmatrix} x^*(t) \\ y^*(t) \end{pmatrix} = \begin{pmatrix} 0{,}1 + 0{,}3t + 0{,}5t^2 \\ 0{,}2 + 0{,}4t + 0{,}6t^2 \end{pmatrix}. \qquad (19)$$

Выберем значения параметров ПД-регулятора (2) как

$$r_0 = 15, \quad r_1 = 20, \quad \ell = 0{,}9g.$$

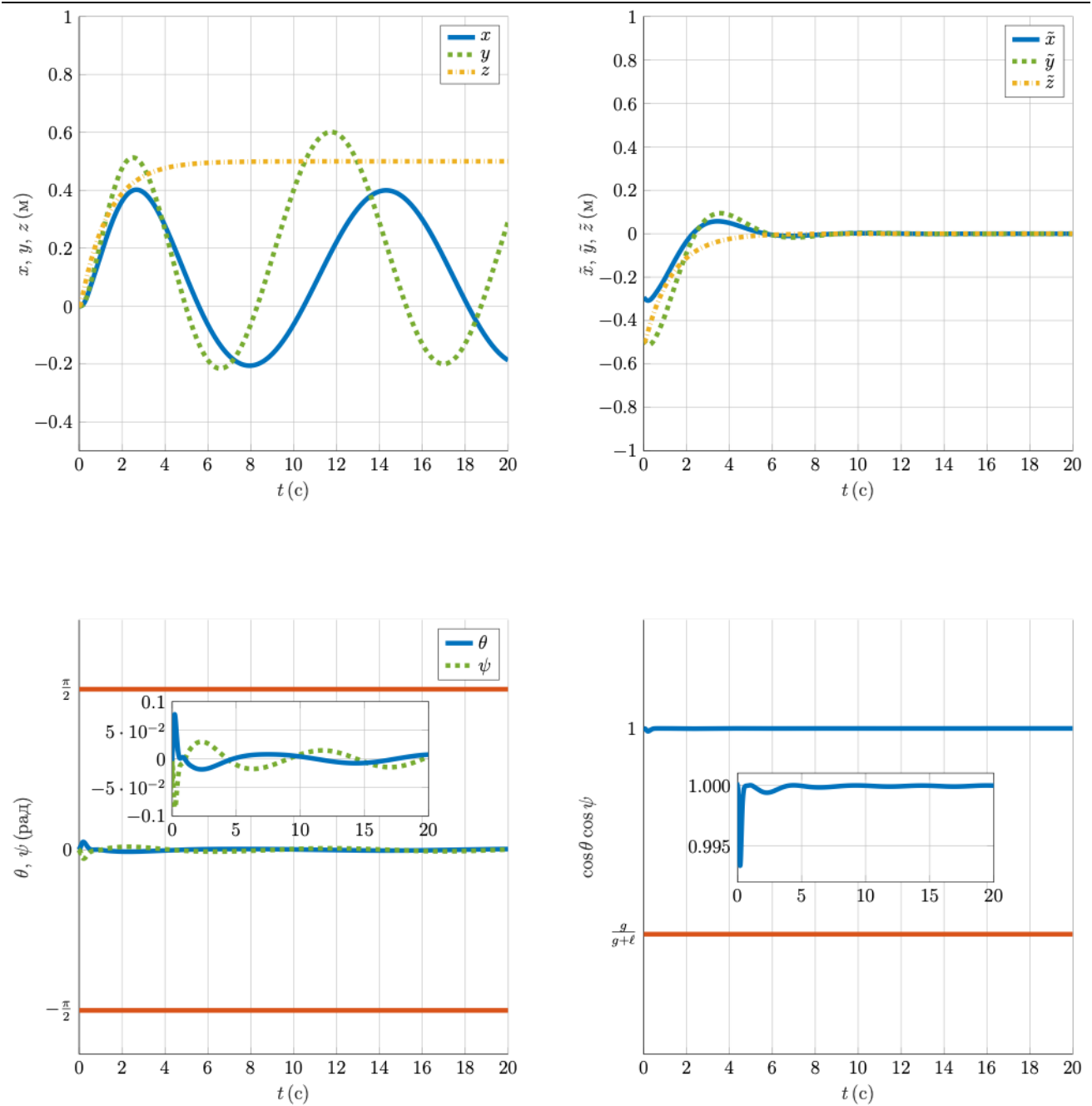

Рис.1. Результаты моделирования с периодическим задающим сигналом

Выберем значения параметров робастного регулятора (16), (17) как

$$\bar{B} = \begin{pmatrix} g & 0 \\ 0 & -g \end{pmatrix}, \quad K = \begin{pmatrix} 100 & 100 & 100 & 10 \end{pmatrix}, \quad \nu = 100, \quad \kappa = 180,$$

а также вычислим значения коэффициентов $a_0, a_1, a_2, a_3, a_4$, следуя процедуре, описанной в работе [9], как

$$\begin{aligned} a_4 &= b_4, \\ a_3 &= b_4 b_3, \\ a_2 &= b_4 b_3 b_2, \\ a_1 &= b_4 b_3 b_2 b_1, \\ a_0 &= b_4 b_3 b_2 b_1 b_0, \end{aligned}$$

где $b_0 = 1$, а коэффициенты $b_1, b_2, b_3, b_4$ вычисляются рекурсивно как

$$\begin{aligned} b_0 &= 1, \\ b_1 &= L_1 b_0 + b_0, \\ b_2 &= \frac{L_2(\sqrt{2} + 2 b_0 b_1) + b_0 b_1}{\beta_{min}}, \\ b_3 &= L_3(\sqrt{3} + 3\beta_{max} b_0 b_1 b_2) + \beta_{max} b_0 b_1 b_2, \\ b_4 &= L_4(2\beta_{max} + 4 b_0 b_1 b_2 b_3) + b_0 b_1 b_2 b_3, \end{aligned}$$

где $L_1 = L_2 = L_3 = L_4 = 1$.

Для слежения за периодическим сигналом (18) зададим значения параметров внутренней модели (11) как

$$F_i = \begin{pmatrix} 0 & 1 & 0 \\ 0 & 0 & 1 \\ -1 & -3 & -3 \end{pmatrix}, \quad G_i = \begin{pmatrix} 0 \\ 0 \\ 1 \end{pmatrix}, \quad \varGamma_i = \begin{pmatrix} 1 & 3 - \rho_i & 3 \end{pmatrix}, \quad i = \{1,2\},$$

где $\rho_1 = 0{,}25$ и $\rho_2 = 0{,}36$.

Для слежения за полиномиальным сигналом (19) в соответствии с Замечанием 2 деактивируем внутреннюю модель, задав

$$\varGamma_i = \begin{pmatrix} 0 & 0 & 0 \end{pmatrix}, \quad i = \{1,2\}.$$

Результаты моделирования приведены на Рисунках 1 и 2. Как видно на графиках, условие $|\cos\theta\cos\psi| \geq \frac{g}{g+\ell}$, необходимое для обеспечения устойчивости вертикального движения по входу-состоянию, выполнено, а также углы крена и тангажа удовлетворяют ограничениям $|\theta| < \pi/2$ и $|\psi| < \pi/2$, при которых замена переменных (6) является обратимой. В обоих моделированиях сигналы ошибок $\tilde{x}, \tilde{y}, \tilde{z}$ сходятся к нулю, а выходные переменные $x, y, z$ стремятся к задающим воздействиям $x^*(t), y^*(t), z^*$.

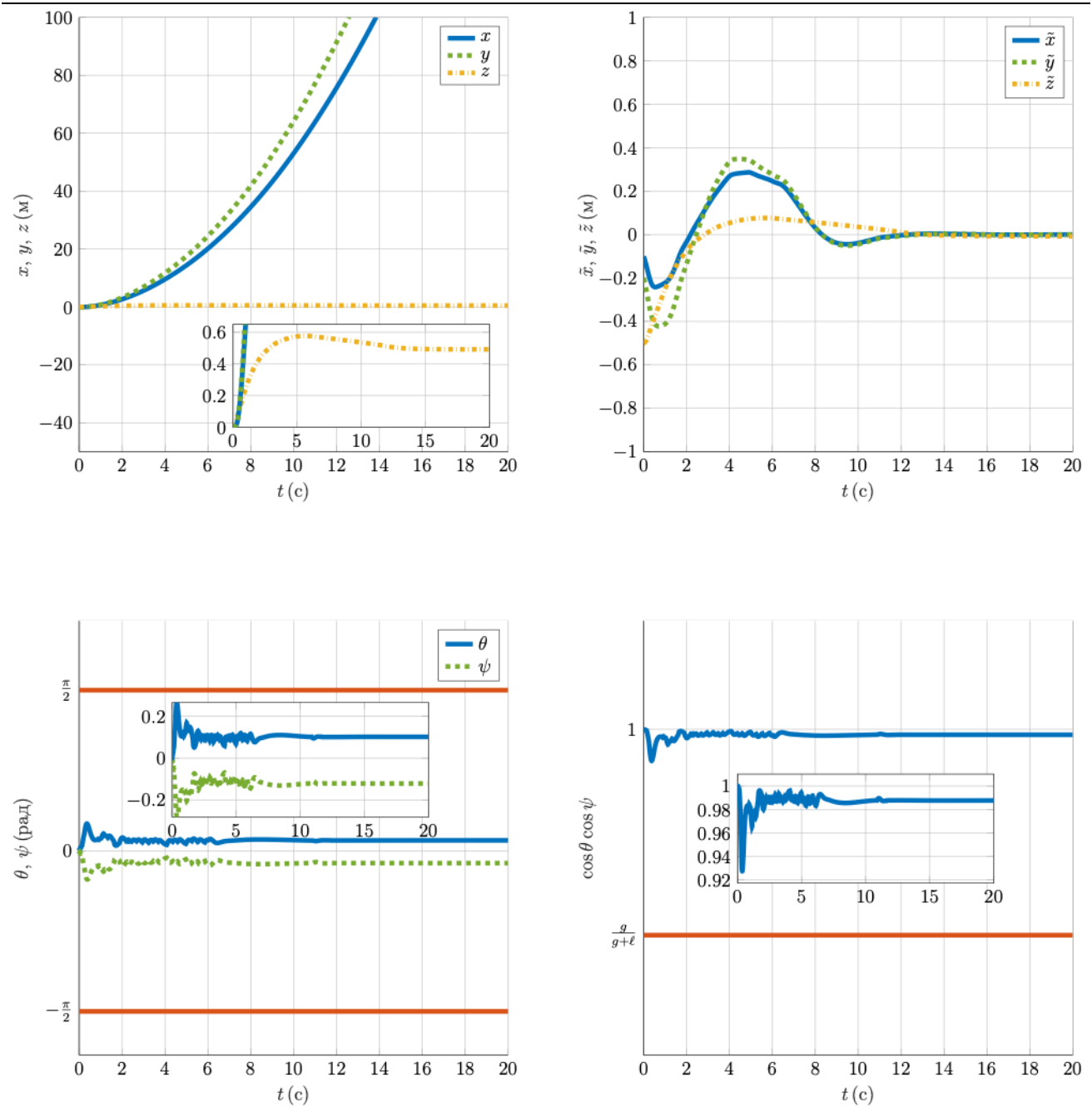

Рис.2. Результаты моделирования с полиномиальным задающим сигналом

**Заключение.** В работе предложено решение задачи траекторного управления моделью квадрокоптера (1) с неизмеряемыми углами тангажа и крена в горизонтальной плоскости по траектории, задаваемой с помощью (4). Предлагаемый подход предусматривает преобразование динамической модели (1) с учетом (2), (4) к модели ошибки в нормальной форме (7), (8), для которой синтезирован закон управления на основе внутренней модели (11) и модифицированного расширенного наблюдателя (16), (17), обеспечивающий достижение полуглобальной асимптотической устойчивости системы и выполнение цели (3). Работоспособность предложенного подхода проиллюстрирована результатами компьютерного моделирования движения квадрокоптера по задающим воздействий в двух формах: периодической (синусоидальной) функции и полиномиальной функции второго порядка.

СПИСОК ЛИТЕРАТУРЫ

GEOMETRY-BASED OUTPUT ROBUST TRACKING CONTROL OF A QUADROTOR

O.I. BORISOV, M.A. KAKANOV, A.Iu. ZHIVITCKII, A.A. PYRKIN

ITMO University, 197101, St. Petersburg, Russia

The paper solves the problem of tracking control of a quadrotor with unmeasurable pitch and roll angles based on the geometric approach with the use of the enhanced extended observer and the internal model. The proposed approach makes it possible to ensure the movement of a quadrotor in a horizontal plane along a trajectory given in the form of a sinusoidal or second-order polynomial function with semiglobal asymptotic convergence of the tracking errors to zero.

Keywords: tracking control, geometric approach, robust control, high-gain observer, internal model, quadrotor.